\documentclass[conference]{IEEEtran}

\usepackage{amsmath,amssymb,amsthm}
\usepackage{epsfig,cite,footnote,authblk,mathtools,enumerate}
\usepackage{color}
\usepackage[table]{xcolor}

\definecolor{bluecolor}{rgb}{0,0.,1.}

\definecolor{redcolor}{rgb}{.7,0.,0.}

\newcommand{\pr}[1]{\left( #1\right)}
\newcommand{\prr}[1]{\left[ #1 \right]}
\newcommand{\es}[1]{\begin{equation}\begin{split}#1\end{split}\end{equation}}
\newcommand{\est}[1]{\begin{equation*}\begin{split}#1\end{split}\end{equation*}}

\newcommand{\R}{\mathbb{R}}

\newcommand{\V}{\mathcal{V}}

\newcommand{\E}{\mathbb{E}}
\newcommand{\A}{\mathcal{A}}

\newcommand{\dd}{\textrm{d}}

\newcommand{\SF}{\text{SF}}
\newcommand{\SNR}{\text{SNR}}
\newcommand{\SIR}{\text{SIR}}

\newcommand{\squeezeup}{\vspace{-1mm}}

\IEEEoverridecommandlockouts

\begin{document}
\bstctlcite{IEEEexample:BSTcontrol}

\title{\huge{LoRa Network Performance Under Ambient Energy Harvesting and Random Transmission Schemes}}

\author{Orestis Georgiou, Constantinos Psomas, Eleni Demarchou and Ioannis Krikidis

\thanks{O. Georgiou, C. Psomas, E. Demarchou and I. Krikidis are with the IRIDA Research Centre for Communication Technologies, Department of Electrical and Computer Engineering, University of Cyprus, Nicosia, Cyprus.}}

\maketitle


\begin{abstract}
LoRa networks have been deployed all over the world and are a major enabling wireless technology for the Internet of Things (IoT). Massive connectivity applications such as smart metering, agriculture, and supply chain \& logistics are most suitable for LoRa deployments due to their long range, low cost, and low power features. Meanwhile, energy harvesting technologies that extract energy from ambient sources have enabled the battery-less operation of many small wireless sensors. This paper studies the merger of these two technologies and mathematically models device and network performance using tools from stochastic geometry and Markov analysis. To that end, we derive the steady-state distribution of the capacitor voltage, the outage probability due to co-spreading factor interference at the LoRa gateway, and propose adaptive charging time schemes in order to mitigate energy outage events.
\end{abstract}


\squeezeup
\section{Introduction \label{sec:intro}}

With more than 25 billion devices being connected today, the Internet of Things (IoT) vision really does start with \textit{connectivity} \cite{li20185g}. 
The main driver for this explosive growth are the immense societal, environmental and industrial benefits enabled through the remote monitoring and control of intelligent processes.
While 5G cellular technologies such as Narrowband-IoT and LTE-Machine Type Communication are progressively being deployed across the world, licensed and unlicensed low-power wide-area network (LPWAN) technologies are also making headway with purpose-built equipment to further support large-scale IoT networks sprawling over vast industrial, public and commercial campuses \cite{lavric2017internet}. 

In this paper, we mathematically model the performance of LoRa${}^\text{TM}$ (Long Range) networks that are powered by ambient energy sources and a small internal capacitor.
To the best of our knowledge, our comprehensive approach  is the first to develop and combine an analytical model of the energy harvesting (EH), the energy storage (ES) and discharge of the capacitor, the LoRa end-device (ED) electrical load model with different asynchronous random access schemes, and a stochastic geometry model of the uplink interference-limited channel.
Each of these components is mathematically tractable, thus enabling the engineering of sub-system parameters. 

The main contributions of this paper are: 
\textit{i)} the formulation of an end-to-end model of ambient EH powered LoRa networks that is amenable to mathematical analysis (see Sec. \ref{sec:system}), 
\textit{ii)} closed form expressions for the uplink performance analysis using a stochastic geometry framework (see eq. \eqref{laplace}),
\textit{iii)} a Markov chain model and steady state distribution for the capacitor voltage enabling the calculation of the ED energy outage probability (see eq. \eqref{Eout2}), and
\textit{iv)} propose adaptive charging time schemes to mitigate energy outage events.

\begin{figure}[t]
\centering
\includegraphics[width=1\columnwidth]{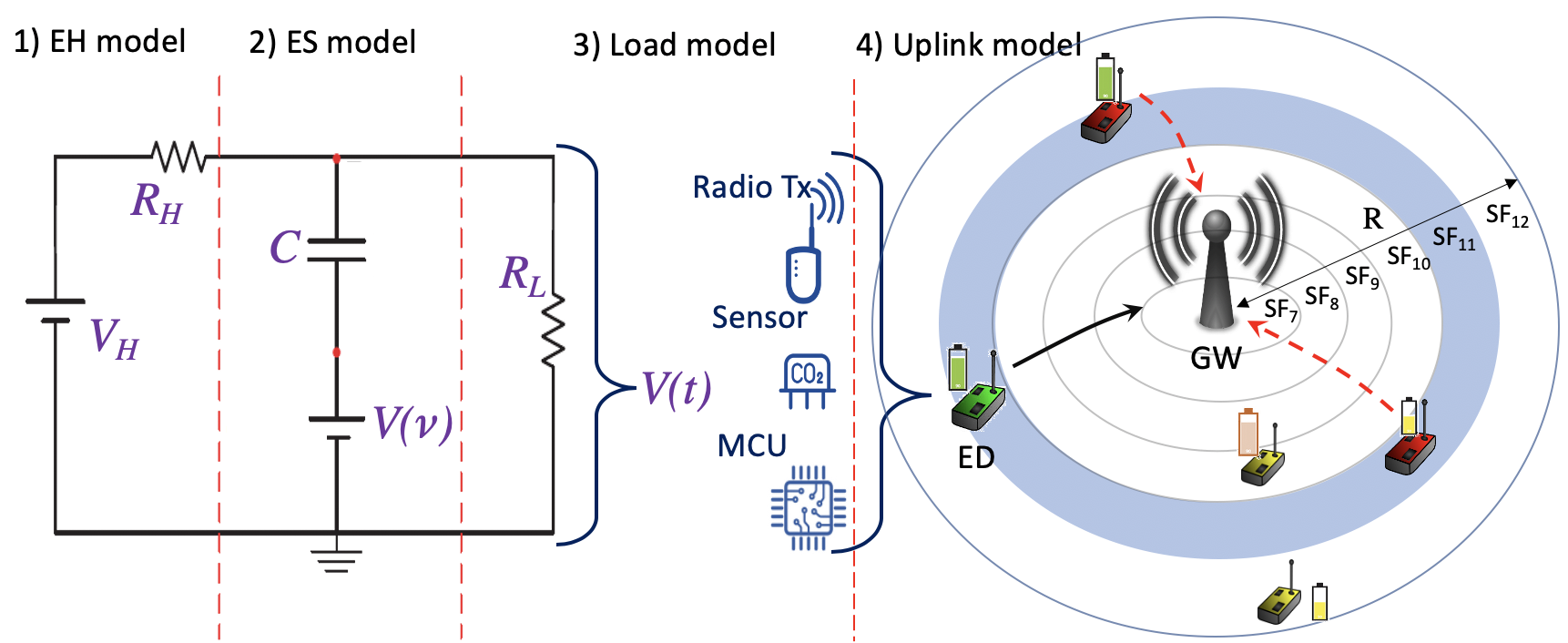}
\caption{ 
Simplified schematic representation of the four sub-models analysed in this paper. 
The EH, ES, and load circuit model is inspired by that presented in \cite{delgadofeasibility} and characterizes the temporal variation of the various electrical components and the capacitor voltage (see Fig. \ref{fig:capacitor}).
The uplink model shows the SF allocation scheme that forms concentric co-SF rings around the LoRa GW and how a transmitting ED (green node) is interfered by other EDs in the same SF ring (red nodes) but not by the ones in other SF rings (yellow).
}
\label{fig:system}
\end{figure}

\squeezeup
\section{Background and Related Works \label{sec:related}}

In general, LPWAN technologies such as SigFox, MIOTY, Weightless, Ingenu and LoRa can only send small blocks of data at a low data rate and are thus better suited for use cases that do not require high bandwidth and are not time-sensitive \cite{khutsoane2017iot}.
LPWANs are thus well positioned to support part of the upcoming 5G ecosystem, particularly due to their massive connectivity, reliability and affordability that makes them well suited for low cost, long range, and low power outdoor applications \cite{raza2017low}.
With a range of multiple kilometers, it is no surprise that LoRa networks, an unlicensed sub-GHz LPWAN technology that uses proprietary chirp spread spectrum (CSS) modulation techniques developed by Semtech, have been deployed in over 150 countries by both private and public operators and have been studied extensively by the academic community. 
Specifically, there have been numerous studies on the performance of LoRA, notably towards improving its capacity, reliability and scalability \cite{georgiou2017low}, e.g., through medium access control (MAC) protocol variations that can reduce the network’s susceptibility to wireless interference and increase capacity and fairness \cite{lim2018spreading,beltramelli2018interference}. 
Moreover, renewable energy sources such as solar, thermoelectric, magnetic or piezoelectric vibrations have also been proposed as means to extend the battery operation of variety of IoT devices \cite{adila2018towards} including LoRa networks but have not been systematically studied \cite{sherazi2018renewable,tjukovs2018experimental}.

LoRa networks that are powered by ambient energy would last longer, require no battery replacement, be cheaper and easier to recycle, and can be better weatherproofed due to the lack of a battery access panel \cite{sherazi2018renewable}.
These are the most important benefits that motivate our current work.
However, this new paradigm is faced with a number of difficulties and challenges. 
For example, low ambient and intermittent EH power sources, small energy storage capacitors, intermittent random access transmissions, and wireless interference can cause frequent power and/or communication outages.

\section{Main Assumptions and System Model \label{sec:system}}

We consider a wireless LoRa network comprising of multiple randomly deployed battery-less ED transceivers, each of which are integrated with some sensors (e.g., temperature, CO/CO2, IMU, proximity, IR, etc.) and with an EH rectification and ES unit. 
The EDs frequently communicate their sensed data in the uplink to a LoRa gateway (GW) that is powered by the mains and connected to a network server (NS) via a reliable back-haul link.
The ED operations (sensing, processing, and wireless transmissions) are solely powered by the ambient EH source that is stored by a capacitor such that a DC output voltage powers the ED operations.
For the sake of mathematical analysis, we assume that the EH rate $P_H$ is constant and equal for all LoRa EDs and therefore do not model any variability in the ES.
This is reasonable for some types of ambient EH sources and sensor deployments considering the difference in time scales of operation, e.g., radio duty cycles occur in milliseconds, while changes in solar irradiance changes occur smoothly over several hours \cite{delgadofeasibility}.
A future in-depth study will focus on temporal and spatial variability of the EH rate.
Here, variability will only be inserted in our model via the wireless channel $|h|^2$, the spatial distribution of EDs $\Phi$, and the random access charging times $\nu$ during which EDs charge their capacitors.

To gain engineering insights we shall deconstruct our system into four sub-models: 
\textit{a)} the EH model, \textit{b)} the ES model, \textit{c)} the ED Load model, and \textit{d)} the uplink communication model, all of which are schematically shown in Fig. \ref{fig:system}. 

\textit{Network Topology:}
We consider a LoRa GW located at the origin of the coordinate system and EDs uniformly located at random in some finite deployment region $\A\subset\R^2$ described through an in-homogeneous Poisson point process (PPP) $\Phi$, with intensity $\lambda>0$ in $\A$, and 0 otherwise.
Each point in the PPP represents an ED. 
For simplicity we assume that $\A$ is a disk of radius $R$ km of area $|\A| = \pi R^2$ km$^2$ containing a total of $N$ EDs, where $N$ is a Poisson distributed random variable with mean $\E[N] = \lambda |\A|$. 
The Euclidean distance from ED $i$ to the GW at the origin is denoted by
$d_i$ km.
Multi GW topologies will not be considered in this paper.

\textit{\textbf{Energy Harvesting Model:}}
The LoRa EDs are completely powered by ambient EH sources. 
For the sake of generality, we do not specify the energy source here and in our simulations assume that the EH rate is constant and equal for all EDs at $P_H$ = 1 mW/cm${}^2$ \cite{delgadofeasibility}.
Such power densities can support many IoT applications and are comparable to that of indoor natural light and vibrational and thermal gradients found in industrial settings \cite{adila2018towards}.
We model the EH circuit as an ideal voltage source and a series resistance (denoted by $V_H$ and $R_H$, respectively).
The EH circuitry is depicted in its simplest form in Fig. \ref{fig:system}.
Note that, the value of $V_H$ (in Volts) is chosen according to the operating voltage of the individual circuit elements, which in this case is determined by the most power hungry component of the ED, i.e., the radio, which according to the SX1272/73 specification is $V_H=3.3 $ V \cite{sornin2015lorawan}. 
Finally, the series resistance $R_H= \frac{V_H^2 }{ P_H} $ limits the current supplied.

\begin{figure}[t]
\centering
\includegraphics[width=\columnwidth]{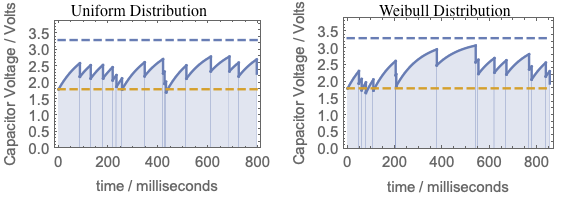}
\caption{ 
Typical time evolution of the capacitor voltage for $\nu\sim \text{Uniform}(a,b)$ on the left sub-figure, and $\nu\sim \text{Weibull}(k,w)$ on the right sub-figure.
Parameters used: $V_H=3.3$ V, $V_i(0)=1.8$ V, $C=10$ mF, $P_H=1$ mW,  $R_L^{(0)}= 600\, \text{k}\Omega$, $R_L^{(1)}= 117 \, \Omega$, $\tau= 204$ ms, and $a=0, b=100, k=1, w=50$ such that $\E[\nu]=50$ seconds for both schemes; a $0.4\%$ duty cycle. 
}
\label{fig:capacitor}
\end{figure}

\textit{\textbf{Energy Storage Model:}}
A capacitor of capacitance $C$ Farads is used to store the energy harvested $E(\nu)= \nu \times P_H$ over a time period of $\nu$. 
Since capacitors store their energy as an electric field rather, they can be recharged over and over again and, unlike batteries, do not lose their capacity to hold a charge.
We model the behaviour of the ES model consists of a succession sub-intervals. 
During the first sub-interval, the capacitor of ED $i$ is being charged for $\nu$ seconds, and then during the second sub-interval it is discharged for $\tau_i$ milliseconds during an uplink transmission to the GW. 
It then charges for another $\nu$ seconds (chosen randomly from a statistical distribution \eqref{pdf}) and then discharges for $\tau_i$, and so on.
The variable $\nu$ is thus both the capacitor charging time and radio inter-transmission time.

We define a \textit{cycle interval} $T_i=\nu + \tau_i$ and model the temporal evolution of the capacitor voltage $V_i(t)$ by the state of each of its load components (see also next sub-section)
\es{
V_i(t)\!= \!\begin{cases}\!
       V_L^{(0)} \big( 1- e^{\frac{-t}{R_L^{(0)} C}}\big) \!+\! V_i(0) e^{\frac{-t}{R_L^{(0)}C}}, \!& \text{if}\ 0\!\leq\! t\!\leq\! \nu  \\
     \! V_L^{(1)} \big( 1- e^{\frac{-(t-\nu)}{R_L^{(1)} C}}\big) \! +\!  V_i(\nu) e^{\frac{-(t-\nu)}{R_L^{(1)}C}}, \!& \text{if}\ \nu \! <\! t \!\leq \! T_i
    \end{cases}
\label{capacitor}
}
where $V_L^{(0)}=\frac{R_L^{(0)}  V_H}{R_H}$ and $V_L^{(1)}=\frac{R_L^{(1)}  V_H}{R_H}$, with the superscript $(0)$ and $(1)$ represents the OFF and ON states of the radio transmitter.
Note that the capacitor voltage $V_i(0)$ at the beginning of a cycle interval equals that at the end of the previous cycle interval, thus introducing memory into the system and dependence on the initial state of the capacitor.
Further, note that during a typical cycle interval the capacitor voltage will fluctuate.
It rises and then drops towards the load voltages of $V_L^{(0)}$ and $V_L^{(1)}$, respectively, within each cycle interval $T_i$ depending on the specific values of $\nu$ and $\tau_i$.
Since $R_L^{(1)} < R_L^{(0)}$, the capacitor discharges faster than it charges.

The behaviour of $V_i(t)$ over several intervals is shown in Fig. \ref{fig:capacitor} for the two random access transmission schemes we study (Weibull and Uniform).
We observe that while $10$ mF Aluminium electrolytic capacitor can be bought off-the-shelf for approx \$1 \cite{delgadofeasibility}, larger capacitors are generally more expensive and may thus not be suitable for some IoT applications.
Also, we observe that the time evolution of the capacitors is very different despite the uplink transmissions following the same duty cycle. 
An energy outage occurs whenever the capacitor voltage drops below the radio operating voltage $\V$.

\textit{\textbf{Load Model:}}
The load of the ED model is made up of the set of its components that consume the stored energy in the capacitor. 
For simplicity, we assume that the majority of the electrical load is consumed by the micro-controller unit (MCU), the radio (R) and the on-board sensors (S). 
Each of these components is characterized by a specific power consumption in each of its states (ON or OFF) that builds up the total load resistance of an ED denoted by $R_L$ in Ohms.
More complex multi-state components can also be included in the model but are deferred for future work. 
Assuming that the MCU and sensors are always ON and that the radio switches ON for  $\tau_i$ milliseconds during uplink transmissions, we can define two main states for the system load resistance
$R_L^{(0)} = \frac{V_H}{I_\text{MCU}+I_\text{S}}$ and $R_L^{(1)} =\frac{V_H}{I_\text{MCU}+I_\text{S} + I_\text{R}}$,
where $I_\text{R} =(P_\text{op}+P_T)/V_H$ milliamperes is the supplied current through the radio transmitter with $P_\text{op}$ being the additional operating power required by the transmission chain, here taken to be $P_\text{op} \approx 6$ dBm \cite{sornin2015lorawan}.
Note that temporal evolution of load resistance $R_L(t)$ as the radio transmitter turns ON/OFF depends on the time on air $\tau_i$ which directly depends on the ED distance $d_i$ to the GW.
For instance, EDs that are further away from the GW will have a higher time on air $\tau_i$ expending more energy according to $\tau_i \times (P_T+P_\text{op})$ per uplink transmission.

\textit{\textbf{Uplink Communication Model:}}

\subsubsection{Radio module}
We consider that each ED is equipped with a Semtech SX1272/73 transceiver module operating at 3.3V within the 868MHz band using the LoRa proprietary CSS modulation scheme.
An omnidirectional antenna transmits at $P_T\!=\!13$ dBm and bandwidth $\text{BW}\!=\! 125$ kHz \cite{sornin2015lorawan}.

\subsubsection{Path Loss and Rayleigh Channel Fading}
To ensure mathematical tractability we assume a path loss attenuation function defined through 
$
g(d_i)= (\psi/4\pi d_i)^\eta
$, 
where $\psi=34.5$ cm is the carrier wavelength, $d_i$ is the Euclidean distance in km between ED $i$ and the central GW, and $\eta\geq 2$ is the path loss exponent.
Also, all wireless links are subjected to both small-scale block fading and large-scale path-loss effects. 
We consider Rayleigh fading (no shadowing effects) such that the channel gain $|h_i|^2$ due to an uplink transmission by ED $i$ is modelled as an exponential random variable of unit mean.

\subsubsection{Random Access Transmission Scheme}
LoRa EDs use the LoRaWAN MAC protocol, which is essentially ALOHA with no collision avoidance provisions \cite{sornin2015lorawan}.  
We thus assume that an ED will attempt to transmit its sensed data at random every $\nu$ seconds chosen from a Weibull or Uniform statistical distribution (WD/UD) with probability density functions (pdf)
\es{
f_{\text{WD}}(x) = \frac{k}{w}\pr{\frac{x}{w}}^{k-1} e^{-(x/w)^k}, \quad f_{\text{UD}}(x) = \frac{1}{b-a}
\label{pdf},
}
where $x, w, k > 0$ and $\quad 0\leq a\leq x \leq b $.
We choose these two distributions as they are very different from each other yet are both characteristic of packet flows in large wireless sensor networks \cite{meng2004characterizing}.
Moreover, the WD interpolates between the exponential distribution and the Rayleigh distribution therefore capturing a range of possible sensor data flow transmissions.
In our analysis, we will investigate how these different transmission schemes impact network performance since more frequent transmissions may require more energy, therefore starving the capacitor and resulting in frequent power outages.

\subsubsection{Interference and Spreading Factor Allocation}
LoRaWAN has provisions for an adaptive data rate (ADR) mechanism using its proprietary CSS modulation scheme for optimizing the allocation of spreading factors (SFs) with respect to the channel conditions. 
Basically, unless otherwise specified by the network operator, an ED will use the ADR mechanism to adapt its SF ranging from 7 to 12 according to a signal-to-noise-ratio (SNR) feedback from the GW.
Thus, EDs with lower SNR margins tend to be located farther away from the GW (due to path loss attenuation) and will therefore opt for a high SF and \textit{vice versa}.
Following this reasoning, we adopt the simplified ring-model proposed in \cite{georgiou2017low} that fixes the SF of each ED depending on its distance to the GW (see Fig. \ref{fig:system}).
The concentric rings that define the SF of a transmitting ED have radii $l_0, l_1, \ldots l_6$ km as described in the last column of Tab. \ref{table:lora}.
Finally, we assume that LoRa chirp signals with different SFs are essentially quasi-orthogonal \cite{croce2018impact}.
We will therefore assume that most of the RF interference at the GW, results from other uplink transmissions originating from EDs located within the same SF ring.
In Fig. \ref{fig:system}, this is illustrated by the red EDs that are in the same SF ring as the green ED.

\begin{table}[t]
\renewcommand{\arraystretch}{1.3}
\caption{LoRa Characteristics of a 25 byte Message at $\text{\text{BW}}\!=\!125$ kHz }
\label{table:lora}
\centering
\scriptsize
\begin{tabular}{| c | c | c | c | c | c |}
\hline
& \bfseries  & \bfseries bit-rate & \bfseries Packet air-  & \bfseries SNR $q_{\text{SF}}$& \bfseries \cellcolor{blue!15}Range \\
$n$ & \bfseries $\text{SF}$ & \bfseries kb/s & \bfseries time $\tau$ ms & \bfseries dBm & \bfseries \cellcolor{blue!15}km \\
\hline\hline
1 & \bfseries 7 & 5.47 & 36.6 & -6 & \cellcolor{blue!15}$l_0\!-\!l_1$ \\ \hline
2 & \bfseries 8 & 3.13 & 64 & -9 & \cellcolor{blue!15}$l_1\! - \!l_2$ \\ \hline
3 & \bfseries 9 & 1.76 & 113 & -12 & \cellcolor{blue!15}$l_2 \!- \!l_3$ \\ \hline
4 & \bfseries 10 & 0.98 & 204 & -15 & \cellcolor{blue!15}$l_3 \!-\!l_4$ \\ \hline
5 & \bfseries 11 & 0.54 & 372 & -17.5 & \cellcolor{blue!15}$l_4 \!-\! l_5$ \\ \hline
6 & \bfseries 12 & 0.29 & 682 &  -20 & \cellcolor{blue!15}$l_5\! -\! l_6$ \\ \hline
\end{tabular}
\end{table}

\subsubsection{Uplink Transmission Time}
The ADR scheme essentially enables EDs to trade-off throughput for coverage range, or robustness, or energy consumption, while keeping a constant bandwidth.
The reduced bit rate thus results in a higher transmit (on-air) time $\tau_i$ milliseconds (see Tab. \ref{table:lora}).
For simplicity, we assume that $\tau_i$ equals the packet payload size divided by the bit-rate, therefore ignoring the time needed to transmit a preamble sequence and the variable coding rate afforded by LoRaWAN \cite{sornin2015lorawan}.
Further, the time-on-air per ED is required by the European Telecommunications Standards Institute (ETSI) to follow a duty cycling policy of up to $1\%$ \cite{raza2017low}, i.e., that $\frac{\tau_i}{ \mathbb{E}[\nu]  +\tau_i} \leq 1\%$, where we can calculate from \eqref{pdf} that $\mathbb{E}[\nu]= m \Gamma(1+\frac{1}{k})$ for WD and $\mathbb{E}[\nu] = \frac{a+b}{2}$ for UD.

\squeezeup
\section{End-Device Performance Analysis \label{sec:analysis}}

There are basically two ways that an uplink transmission from an ED $i$, located $d_i$ km from the LoRa GW may fail:
\begin{enumerate}[i)]
\item{The capacitor voltage of ED $i$ falls below the operating threshold $\mathcal{V}$ and is unable to complete its transmission.}
\item{The GW cannot decode the uplink radio communication from ED $i$ because the received signal to noise ratio (SNR) or the signal to interference ratio (SIR) at the GW are less than the thresholds $q_{\SF_i}$ and $\wp$ respectively.}
\end{enumerate}
Combined, these two conditions define the per-device uplink reliability and thus in effect that of the whole LoRa network.
The outage probability due to the energy condition i) is $\bar{\mathcal{E}}(d_i) = 1 - \mathcal{E}(d_i)$, and the communication outage condition ii) is $\bar{\mathcal{C}}(d_i) = 1 - \mathcal{C}(d_i)$, such that the overall probability of a successful uplink transmission is given by
$
\mathcal{Q}(d_i) = \mathcal{E}(d_i)\mathcal{C}(d_i).
$

\textit{\textbf{Communication Outage:}}
In this subsection we study the communication outage $\bar{\mathcal{C}}(d_i)$ of LoRa EDs and show how it is connected to the energy outage $\bar{\mathcal{E}}(d_i)$.
Unlike other communications systems, e.g., cellular, where interference is treated as shot-noise and performance is measured by the signal-to-interference-plus-noise-ratio (SINR), LoRa uplink transmissions have a dual requirement that separates SNR and SIR. 
This is due to the CSS modulation scheme which allows GWs to decode multiple received transmissions, as long as their relative signal strengths are sufficiently distinct \cite{beltramelli2018interference}.
Specifically, we say that an uplink transmission by ED $i$ is successful if the condition 
$
\pr{\SNR_i \geq q_{\SF_i}  } \bigcap \pr{\SIR_i \geq \wp }$
is met, where $q_{\SF_i}$ is the SNR condition taken from Tab. \ref{table:lora}, and $\wp = 1\text{ dB }= 1.259$ \cite{beltramelli2018interference} where we have defined 
$\SNR_i = \frac{P_T |h_i|^2 g(d_i)}{\mathcal{N}}
\label{SNR}$
and
$
\SIR_i = \frac{P_T |h_i|^2 g(d_i)}{ \mathcal{I}_i}$ 
and $\mathcal{I}_i = \sum_{k\not=i} \chi_{ik} P_T |h_k|^2 g(d_k)$ is the total co-SF interference at the GW met by an uplink transmission of ED $i$ due to EDs $k\not=0$ with $\chi_{ik}(t)$ being 1 if EDs $i$ and $k$ are both transmitting using the same SF and at the same instant in time, and 0 otherwise, the average of which is the packet collision probability $\mathbb{E}[\chi_{ik}] = p_i$.
Putting the above together, we have that the connection probability is
$
\mathcal{C}(d_i)=\mathbb{P}\prr{ \pr{\SNR_i \geq q_{\SF_i}  } \bigcap \pr{\SIR_i \geq \wp }  } \geq \mathbb{P}\prr{ \SNR_i \geq q_{\SF_i}} \times \mathbb{P} \prr{\SIR_i \geq \wp } $,
where we have assumed independence between these events, thus resulting in the lower bound probability inequality (see \cite[eq.(5)]{lim2018spreading}).
Note that an upper bound to $\mathcal{C}(d_i)$ can be obtained by using $\max(x,y)=(x+y+|x-y|)/2\geq (x+y)/2$ and  $\mathcal{C}(d_i)=\mathbb{P}\big[|h_i|^2\geq  \max \big(\frac{\mathcal{N} q_{\SF_i}}{P_T g(d_{i})}, \frac{w\mathcal{I}_i}{P_T g(d_{i})} \big) \big| \, d_i \big]$.

\subsubsection{SNR Analysis}
Following from the definition of SNR and since we assumed a Rayleigh fading channel model we calculate
$
\mathbb{P}\prr{ \SNR_i \geq q_{\SF_i}}  = \mathbb{P}\Big[ |h_i|^2 \geq \frac{\mathcal{N} q_{\SF_i}}{P_T g(d_i)} \Big]  = e^{-\frac{\mathcal{N} q_{\SF_i}}{P_T g(d_i)}}
$.
Note that $q_{\SF_i}$ is a piece-wise function of $d_i$, such that $q_\SF (d_i) = 10
^{-q/10}$ with $q = \{ 6, 9, 12, 15, 17.5, 20\}$ if $d_i$ falls in the corresponding range (see Tab. \ref{table:lora}). 
This distance dependent sensitivity of $q_\SF (d_i) $ results in $\mathbb{P}\prr{ \SNR_i \geq q_{\SF_i}} $ having a non-smooth, saw-tooth-like profile (see \cite[Fig. 2.a]{georgiou2017low}).

\subsubsection{SIR Analysis}
Following the definition of SIR and taking a stochastic geometry approach \cite{haenggi2012stochastic} to calculating averages of PPPs we have that for a Rayleigh fading channel
$
\mathbb{P} \prr{   \text{SIR}_{i} \geq \wp  } \!=\! \mathbb{E}_{\mathcal{I}_i} \Big[ \mathbb{P} \Big[  |h_i|^2 \geq \frac{\wp \mathcal{I}_i}{P_T g(d_{i})}  \Big| \mathcal{I}_i \Big] \!=\! \mathbb{E}_{\mathcal{I}_i} \Big[  e^{-\frac{ \wp \mathcal{I}_i}{P_T g(d_i)}  }\Big]
$,
where we identify the last term as the Laplace transform of the random variable $\mathcal{I}_i$ evaluated
at $ \frac{\wp }{P_T g(d_{i})}$ conditioned on the distance of the ED $d_i$ km from the LoRa GW.
Expanding the Laplace transform we get
$
\mathbb{E}_{\mathcal{I}_i} \Big[  e^{-\frac{ \wp \mathcal{I}_i}{P_T g(d_i)}  }\Big]
\!=\! \mathbb{E}_{d_{k}} \Big[ \prod_{k\not=i} \mathbb{E}_{|h_{k}|^2}  \Big[ e^{-   \wp \chi_{ik} |h_{k}|^2 \frac{d_{i}^\eta}{d_{k}^\eta} } \Big] \Big] 
 \!=\!  \mathbb{E}_{d_{k}}  \Big[  \prod_{k\not=i}  \Big( 1 + \wp  \chi_{ik}  \frac{d_{i}^\eta}{d_{k}^\eta} \Big)^{-1} \Big]$,
since the channel gains $|h_{k}|^2$ are independent exponential random variables and we used the fact $\mathbb{E}_{|h|^2} [e^{-x |h|^2}] = \int_0^\infty e^{-z(x+1)} \dd z= \frac{1}{1+x} $  \cite{haenggi2012stochastic}.
Now, assuming that the EDs interfering with the uplink transmission of ED $i$ are distributed according to a thinned PPP $\Phi_i\subset \Phi $ with intensity $p_i\lambda <\lambda$ enables us to absorb the $\chi_{ik}$ term and use the probability generating function (PGF) of an inhomogeneous PPP \cite{haenggi2012stochastic}
to finally arrive at a closed form 
\es{
&\mathbb{P} \prr{   \text{SIR}_{i} \geq \wp  } = \exp\Big(-2 \pi p_i\lambda  \int_{l_{n-1}}^{l_{n}} \frac{\wp  \frac{d_{i}^\eta}{d_{k}^\eta}}{1+\wp  \frac{d_{i}^\eta}{d_{k}^\eta}} d_{k} \dd d_{k} \Big)
 \\
 &= \exp\Big(\!\!- 2 \pi p_i  \lambda  \wp  d_i^\eta  \Big[ \frac{ l_n^{2-\eta}}{2-\eta} F_{21}(l_n) -  \frac{ l_{n-1}^{2-\eta}}{2-\eta} F_{21}(l_{n-1})   \Big]    \Big) 
\label{laplace},
}
where we defined $F_{21}(x)={}_2F_1[1,\frac{2}{\eta},1+\frac{2}{\eta}, -\frac{x^\eta}{\wp d_i^\eta}]$ to save space;  ${}_2F_1$ is the Gauss Hypergeometric function that can be computed using standard software packages.
Note that the integration limits effectively only consider co-SF interference. 

\begin{figure}[t]
\centering
\includegraphics[width=\columnwidth]{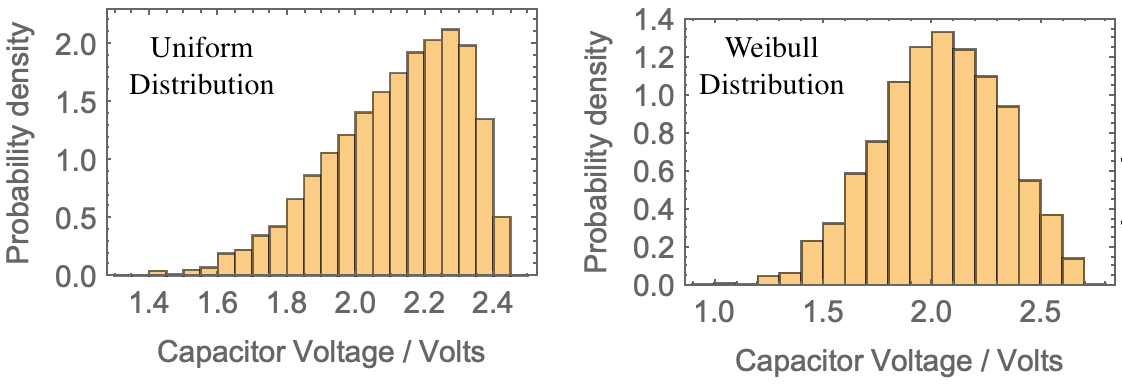}
\caption{ 
Probability density distribution of capacitor voltages following the same transmission schemes and parameters used in Fig. \ref{fig:capacitor}. 
}
\label{fig:steady}
\end{figure}

The value of $p_i\in[0,1]$  in \eqref{laplace} plays an important role in the performance of battery-less EDs since it represents the fraction of co-SF EDs that are transmitting at the same time and SF as ED $i$. 
It therefore depends on the distribution of the charging times $\nu$, the time on air $\tau_i$ for EDs $i$ and $k$, and their energy outage probability.
More specifically it can be modeled as the product of the probability that ED $k$ is not in energy outage multiplied by the per-SF duty cycle $p_i= \mathcal{E}(d_i)\times \mathbb{E}[\frac{\tau_i}{\nu +\tau_i}]$ since $\mathcal{E}(d_k)=\mathcal{E}(d_i)$. 
Observe that a high probability of transmission $\mathcal{E}(d_i)$ will indirectly have a deteriorating effect towards its up-link reception at the GW; an interesting trade-off in the form of 
$\mathcal{Q}(d_i) = \mathcal{E}(d_i)\times \mathcal{C}(d_i)\sim \mathcal{E}(d_i) \times  e^{- \mathcal{E}(d_i)}$ that has not been previously studied.
Equation \eqref{laplace} is the first main result of this paper since it relates communication outage to energy outage.

\textit{\textbf{Energy Outage:}}
Using the model for the capacitor voltage, we define the outage probability $\bar{\mathcal{E}}(d_i)$ of ED $i$ as the probability that the capacitor voltage at the end of a cycle interval is less than the required operating one which according to the SX1272/73 specifications is $\V=1.8$ V. 
In order to make progress, we investigate the memory effect of the capacitor voltage during discrete incremental steps of length $\nu$ and $\tau_i$
\es{
&v_1 = V_i(\nu)=  V_L^{(0)}  +  (v_0 - V_L^{(0)})e^{\frac{-\nu}{R_L^{(0)} C}} \\
&v_2= V_i(T_i) =  V_L^{(1)} + (v_1 -  V_L^{(1)})e^{\frac{-\tau_i}{R_L^{(1)} C}}
}
which generalizes to 
\es{
&v_{2n+1}= V_i(\sum_{j=1}^n T_i +\nu) = V_L^{(0)}  + (v_{2n} - V_L^{(0)}) e^{\frac{-\nu}{R_L^{(0)} C}}\\
&v_{2n+2}= V_i(\sum_{j=1}^{n+1} T_i) =  V_L^{(1)} + (v_{2n+1} -  V_L^{(1)})e^{\frac{-\tau_i}{R_L^{(1)} C}}
\label{regressive},
}
where $v_0= V_i(0)$. 
Note that the last line of \eqref{regressive} describes a first order multiplicative auto-regressive stochastic process of the form
$
v_{2n +2} = c_1+ c_2  X_z (v_{2n} - c_3)
$
where $X_z=e^{\frac{-\nu}{R_L^{(0)} C}}$ is a function of the random variable $\nu$ for $z=\{\text{WD, UD}\}$, and therefore a random variable itself. 
We can thus define
\es{
\bar{\mathcal{E}}(d_i) \!=\! \lim_{n\to\infty}\mathbb{P}[v_{2n+1} \leq \mathcal{V}] \!=\! \lim_{n\to\infty} \!\Big\langle \frac{1}{n}\!\sum_{j=1}^{n}\! H(\V\!-\!v_{2j}) \Big\rangle
\label{Eout}
}
where $H(x)$ is the Heaviside step function such that $H(x)=1$ if $x\geq 0$, and 0 otherwise, and the angled brackets represent an average over initial conditions $V_i(0) \in[V_L^{(1)},V_L^{(0)} ]$.

Fig. \ref{fig:steady} shows the distribution of the capacitor states at the end of cycle intervals $v_{2n}$, averaged in a Monte-Carlo fashion over many initial conditions $V_i(0)$ for the two transmission schemes WD/UD under consideration. 
We observe that the distributions appear unimodal and skewed to the right with a well defined mean.
Moreover, by running multiple simulations using different seeds for $V_i(0)$ we have observed that the two distributions are insensitive to the initial capacitor voltage state $V_i(0)$ for large $n$, thus suggesting that they converge to a steady state and, therefore, the probability distributions of $v_{2n}$ and $v_{2n+2}$ can be assumed identically and independently distributed at large  $n$ such that
$
\mathbb{P}[v_{2n+1} \leq \mathcal{V}] = \mathbb{P}[c_1+ c_2  X_z (v_{2n} - c_3)\leq \mathcal{V}] =\mathbb{P}[v_{2n} \leq \mathcal{V}]
$,
where we define the positive constants $c_1=  V_L^{(1)} +  (V_L^{(0)} - V_L^{(1)})e^{\frac{-\tau_i}{R_L^{(1)} C}}  $, $c_2= e^{\frac{-\tau_i}{R_L^{(1)} C}}$, and $c_3 =  V_L^{(0)}$, for brevity.

An estimator of the mean capacitor voltage can be obtained by solving $v_{2n}=c_1+c_2 \mathbb{E}[X_z](v_{2n}-c_3)$ for $v_{2n}$, to get $\mathbb{E}[v_{2n}] \approx \frac{c_2 c_3 \mathbb{E}[X_z] - c_1 }{c_2 \mathbb{E}[X_z] -1}$, where $\mathbb{E}[X_z]= \int f_z(x) e^{\frac{-x}{R_L^{(0)}C}} \dd x$ that can be easily calculated from \eqref{pdf}.
Recall that $c_1$ and $c_2$ depend on $\tau_i$ and hence the distance $d_i$.
This is the second main result of this paper since it enables one to calculate $\mathbb{E}[v_{2n}]$ for different co-SF rings and test if $\mathbb{E}[v_{2n}]\geq \V$.
In order to calculate $\mathcal{E}(d_i)$ we must retrieve the full distribution of $v_{2n}$. 

\textit{\textbf{Markov Chain Model and Steady State Distribution:}}
To model of the capacitor voltage steady state we start by discretizing all possible states $v_{2n}^{(k)}$ into $M$ bins of equal size $\delta= \frac{V_L^{(0)} -V_L^{(1)}}{M}$, with $k\in[1,M]$.
Then, we calculate the pdf of $X_z=e^{\frac{-\nu}{R_L^{(0)} C}}$ by recognising that 
$\mathbb{P}[X_z \leq x] = \mathbb{P}[\nu> -R_L^{(0)} C \ln x]= 1- F_{z}(-R_L^{(0)} C \ln x)$ for $z=\{\text{WD},\text{UD} \}$ and then differentiate with respect to $x$ to obtain $f_{X_z}(x) = \frac{R_L^{(0)} C}{x}  f_{z}(-R_L^{(0)} C \ln x)  $.
Recall that $f_{z}(x)$ was given in \eqref{pdf} and that $X_z$ is supported on $\big[e^{-\frac{b}{R_L^{(0)} C}},e^{-\frac{a}{R_L^{(0)} C}}\big]$ for UD and the positive real axis for WD.
Then, we seek to construct an $M\times M$ transition matrix $\mathbf{S}$ with entries $S_{ij}$ that describes the transition probability between descritized capacitor states 
\est{
S_{ij}&= \mathbb{P}[v_{2n}^{(i)} \to v_{2n+2}^{(j)}] = \mathbb{P} [v_{2n +2}^{(j)} = c_1+ c_2  X_z (v_{2n}^{(i)} - c_3)] \\
&= \mathbb{P} \Big[ X_z = \frac{v_{2n+2}^{(j)}-c_1}{c_2( v_{2n}^{(i)} - c_3)}\Big] 
= f_{X_z} \Big( \frac{v_{2n+2}^{(j)}-c_1}{c_2( v_{2n}^{(i)} - c_3)}\Big).
}
It is important to set $S_{ij}$ to zero whenever $\frac{v_{2n+2}^{(j)}-c_1}{c_2( v_{2n}^{(i)} - c_3)}$ falls outside of the support of $X_z$ as that would correspond to an impossible transition.
As a final preparatory step, we then row-normalize $\mathbf{S}$ such that each row sums to 1, thus making our new matrix $\hat{\mathbf{S}}$ a right stochastic matrix.
By taking the transpose $\hat{\mathbf{S}}^\text{T}$ and calculating the leading right eigenvector $\mathbf{u}=\{u_1,u_2\ldots u_M \}$ with eigenvalue 1 using standard computational packages we finally arrive at the stationary distribution of $v_{2n}$ given by the  vector $\hat{\mathbf{u}}=\mathbf{u}/ |\mathbf{u}|_{1}$, where $|\mathbf{u}|_1$ is the $L^1$ norm of the vector $\mathbf{u}$.
Notice that the limit $M\to \infty$ corresponds to the continuous capacitor voltage case.
Finally, following from the aforementioned formulation, we can write
\es{
\bar{\mathcal{E}}(d_i) = \lim_{n\to\infty} \mathbb{P}[v_{2n} \leq \mathcal{V}]= \lim_{M\to \infty} \sum_{k=1}^{\kappa} \hat{u}_k
\label{Eout2},
}
with $\kappa$ being the largest value of $k$ that satisfies $v_{2n}^{(k)} \leq  \mathcal{V}$. 
Similarly, the pdf of $v_{2n}$ is given by $\mathbb{P}[v_{2n}= v_{2n}^{(k)}] = \hat{u}_k/\delta$ in the limits $n,M\to\infty$, as shown in Fig. \ref{fig:pdf}.

\begin{figure}[t]
\centering
\includegraphics[width=0.95\columnwidth]{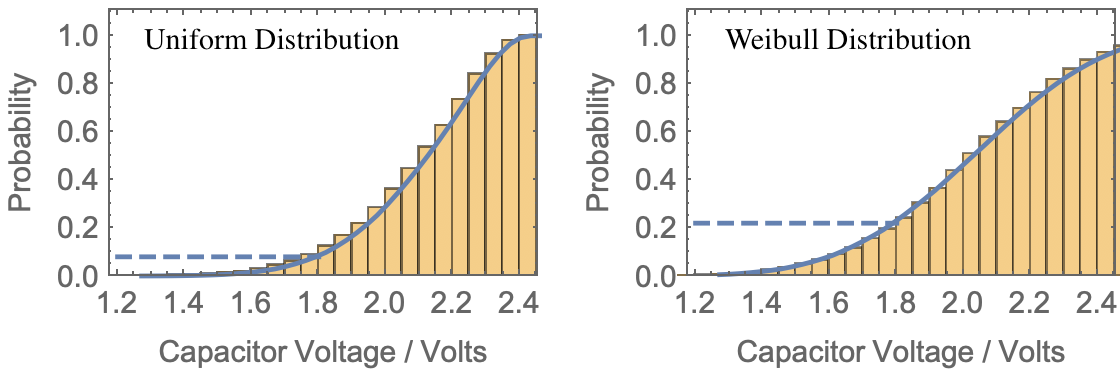}
\caption{ 
Cumulative distribution function of capacitor voltages following the same transmission schemes and parameters used in Fig. \ref{fig:capacitor}. The blue curves correspond the predicted calculations obtained from \eqref{Eout2}, while the dashed horizontal lines shows the probability of energy outage when $\mathcal{V}=1.8$ Volts.
}
\label{fig:markov}
\end{figure}

\begin{figure}[t]
\centering
\includegraphics[width=0.95\columnwidth]{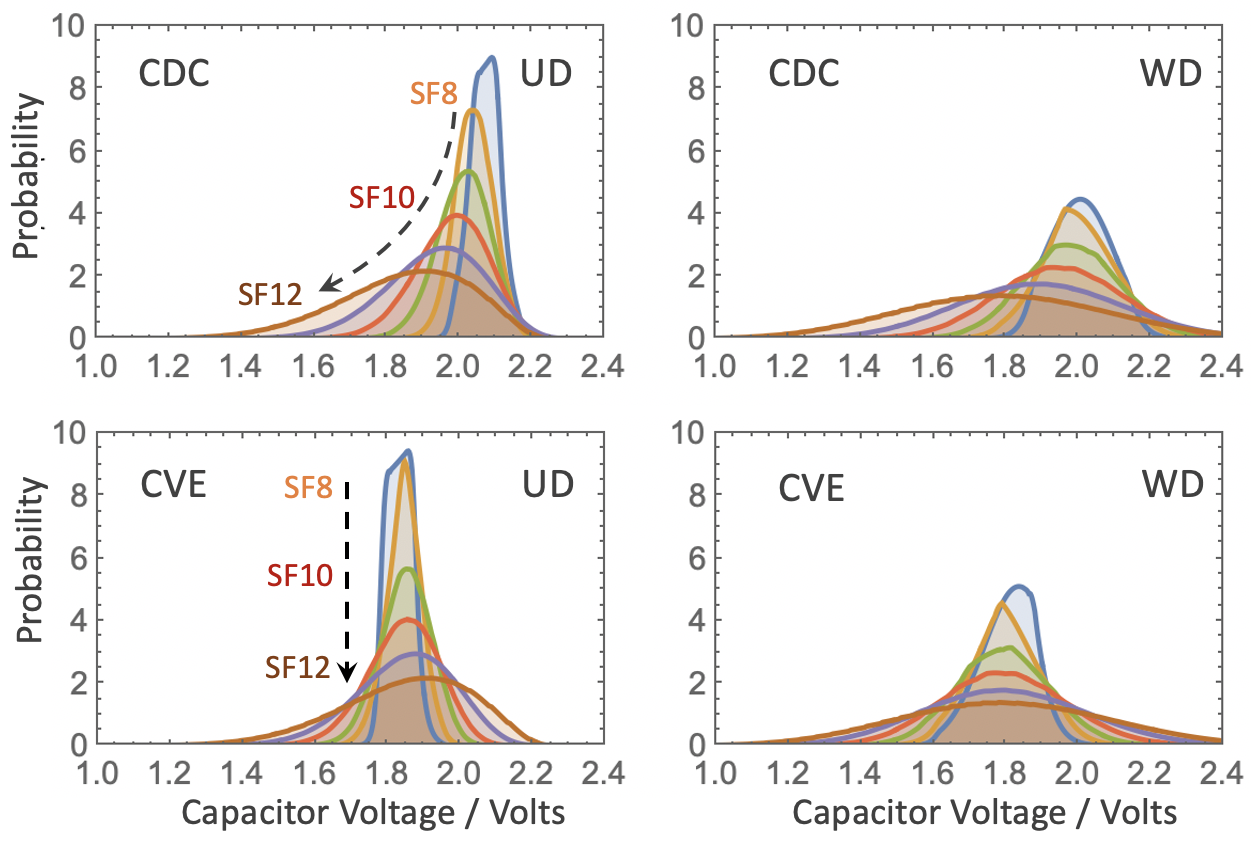}
\caption{Plots of the capacitor voltage pdf \eqref{Eout2} for the two transmission schemes (UD/WD) and the two ACT schemes (CDC/CVE), using $C=40$ mF, $\theta=150, \vartheta= 1$, $b=2\theta,w=\theta$, and all other parameters as in Fig. \ref{fig:capacitor}.
}
\label{fig:pdf}
\end{figure}

Equation \eqref{Eout2} is the third main result of this paper as it enables the calculation of $\mathcal{E}(d_i)$, which can then be used for calculating $\mathcal{C}(d_i)$ and $\mathcal{Q}(d_i)$ as demonstrated earlier in this section.
Fig. \ref{fig:markov} compares the Markov chain derived cumulative distribution function (cdf) of $v_{2n}$ (calculated using \eqref{Eout2}), to the numerically simulated 
\eqref{Eout} for both WD and UD transmission schemes.
Notice that while the two distributions for the charging times $\nu$ used have the same mean charging time $\mathbb{E}[\nu]=50$ seconds and duty cylce of $0.4\%$, the WD shown in the right sub-figure of Fig. \ref{fig:markov} is much more likely to be in energy outage than the UD.
For instance, we have that $ \mathbb{P}[v_{2n} \leq \V]$ is 8\% for UD and 22\% for WD, thus implying that UD outperforms WD in this instance and that care should be taken when choosing a random access transmission scheme.

\textit{\textbf{Adaptive Charging Times:}}
As stated previously, the LoRa ADR scheme enables EDs to trade-off throughput or energy consumption.
This can have a detrimental effect on ED performance if powered by ambient EH.
For example, using eq. \eqref{Eout2}, we can calculate the energy outage probability $\bar{\mathcal{E}}(d_i)$ for the parameters which are used in Fig. \ref{fig:capacitor} with a constant mean charging time $\mathbb{E}[\nu]=50$ seconds but for different time on air values $\tau_i= \{ 36.6, 64, 113,204, 372,682 \}$ milliseconds corresponding to the different SFs in Tab. \ref{table:lora}, to get that $\bar{\mathcal{E}}(d_i) = \{ 0\%, 0\%, 0\%, 8\%, 81\%, 100\% \}$ for UD, and 
$\bar{\mathcal{E}}(d_i) = \{0\%, 0\%, 0.2\%, 22\%, 80\%, 100\% \}$ for WD, respectively, thus suggesting that SFs 11 and 12 will suffer severe outages due to insufficient capacitor voltage.
This could be mitigated by using a larger (expensive) capacitor $C$ and/or longer charging times $\nu$.
We thus propose two adaptive charging time (ACT) schemes that vary the mean charging time $\mathbb{E}[\nu_i]$ of ED $i$ according to the allocated SF.
\emph{1) Constant Duty Cycle (CDC)}
In this scheme, we chose the charging time parameters $(a, b)$ and $(k,w)$ for each SF such that
$\mathbb{E}[\nu_i]= \theta \tau_i$, where $\frac{1}{1+\theta}$ is the uplink duty cycle.
\emph{2) Capacitor Voltage Equalization (CVE)}
In this scheme, we chose the charging time parameters $(a, b)$ and $(k,w)$ for each SF such that mean capacitor voltage 
$\mathbb{E}[v_{2n}]= \vartheta \V$ is the same for each SF, or  that $\mathbb{E}[X_z]= \frac{\vartheta - c_1}{c_2(\vartheta-c_3)}$.

It is seen in Fig. \ref{fig:pdf} that the CDC scheme results in EDs at lower SFs to over-perform in terms of energy outage, thus implying that the transmission duty cycle is unnecessarily too low. 
In contrast EDs using the CVE scheme will on average perform the same, regardless of where they are located in the network.
We also observe that WD distributions are more spread out thus implying more energy outages.
The most robust and efficient scheme appears to be a UD for uplink transmissions and CVE for the ACT.

\squeezeup
\section{Conclusion \label{sec:conclusion}}

We have investigated the performance of battery-less LoRa networks powered by ambient EH sources under random transmission schemes.
Specifically, we have developed a comprehensive and mathematically tractable model for each of the system sub-components (see Fig. \ref{fig:system}) and obtained expressions for the energy and communication outage probabilities by leveraging tools from stochastic geometry and Markov chain analysis.
Importantly, we have demonstrated that the adaptive data rate of LoRa networks can cause energy outages when using higher SFs, and propose adaptive charging time schemes as an effective countermeasure.
We expect that our results will qualitatively remain unchanged if some of our assumptions change, however a full parameter investigation would certainly help in capturing this.
We conclude that battery-less LoRa networks are a feasible solution and could form the next generation of energy-neutral wireless sensor networks for massive connectivity and smart IoT applications.

\squeezeup
\section*{Acknowledgements}
We acknowledge funding from the EU's H2020 research and innovation programme under the Marie Sk{\l}odowska-Curie  project 787180 (NEWSENs), the European Research Council project 819819, the EU Regional Development Fund and the Republic of Cyprus through the Research and Innovation Foundation projects INFRASTRUCTURES/1216/0017 (IRIDA) and POST-DOC/0916/0256 (IMPULSE).

\squeezeup
\bibliographystyle{ieeetran}
\bibliography{mybib}

\end{document}